\documentclass[preprint]{ptephy_om}

\preprintnumber{OU-HET-1173}

\numberwithin{equation}{section}

\usepackage{hyperref}

\title{Higher-group structure in lattice Abelian gauge theory under instanton-sum modification}
\author{Naoto Kan\orcid{0000-0001-9204-8690}\thanks{naotokan000@gmail.com}}
\author{Okuto Morikawa\orcid{0000-0002-0044-4491}\thanks{o-morikawa@het.phys.sci.osaka-u.ac.jp}}
\author{Yuta Nagoya\thanks{y\_nagoya@het.phys.sci.osaka-u.ac.jp}}
\author{Hiroki Wada\thanks{hwada@het.phys.sci.osaka-u.ac.jp}}
\affil{Department of Physics, Osaka University,
Toyonaka, Osaka 560-0043, Japan}

\begin{document}
\begin{abstract}
 We consider the $U(1)$ gauge theory on a four-dimensional torus,
 where the instanton number is restricted to an integral multiple of $p$.
 This theory possesses the nontrivial higher-group structure,
 which can be regarded as a generalization of the Green--Schwarz mechanism,
 between $\mathbb{Z}_q$ $1$-form and $\mathbb{Z}_{p q}$ $3$-form symmetries.
 Here, $\mathbb{Z}_q$ is a subgroup of the center of~$U(1)$.
 Following the recent study of the lattice construction
 of the $U(1)/\mathbb{Z}_q$ principal bundle,
 we examine how such a structure is realized on the basis of lattice regularization.
\end{abstract}
\subjectindex{B01, B06, B31, B35}
\maketitle

\section{Introduction}
The concept of symmetry is a powerful tool
for studying quantum field theories (QFTs).
It is quite often very useful to consider the 't~Hooft anomaly~\cite{tHooft:1979rat},
which is a quantum anomaly when a global symmetry in the theory is gauged,
and is notably invariant under the renormalization group flow.
Thus, the 't~Hooft anomaly matching condition between anomalies in low and high energies
restricts the phase structure of strongly coupled field theories.
In the last decade, generalized symmetries have been vigorously
researched~\cite{Gaiotto:2014kfa} 
(see also Refs.~\cite{McGreevy:2022oyu,Cordova:2022ruw} for reviews).
The higher-form symmetries, a type of generalized global symmetry,
act not only on zero-dimensional objects
but also on higher-dimensional objects in the theory,
and then its 't~Hooft anomaly matching condition provides us
with the nontrivial restriction on low-energy dynamics of physical systems~\cite{Gaiotto:2017yup}.

Such symmetries in a quantum theory may not be
a naive direct product of those symmetry groups.
In general, some symmetries may form
the so-called higher-group structure as a categorical construction~\cite{Kapustin:2013qsa,Kapustin:2013uxa}.
In fact, examples of QFTs have been reported
for both continuous and discrete symmetries~\cite{Tachikawa:2017gyf,Cordova:2018cvg,Benini:2018reh,Hidaka:2020iaz,Cordova:2020tij,Hidaka:2021kkf,Nakajima:2022feg}.
For example, it is possible for some theories that there is
a $0$-form symmetry $G$, and a $1$-form symmetry $H$ can be combined
into a $2$-group structure.
Then the $2$-form gauge field of the $1$-form symmetry $H$ must transform
under not only the usual $1$-form gauge symmetry but also the $0$-form symmetry.
When both symmetries $G$ and $H$ are continuous,
this is nothing but the Green--Schwarz mechanism~\cite{Green:1984sg}.
Even in the case of the discrete symmetries,
the higher-group structure can be interpreted in the language of symmetry defects~\cite{Tachikawa:2017gyf,Benini:2018reh}.
The point is that when the symmetries construct the higher-group structure,
we can no longer consider those individually.

It was recently pointed out by Tanizaki and \"Unsal~\cite{Tanizaki:2019rbk} that
the pure $SU(N)$ Yang--Mills (YM) theory with modified instanton sum
is equipped with higher-group symmetry.
Modification of the instanton sum was originally discussed in Refs.~\cite{Pantev:2005rh,Pantev:2005wj,Pantev:2005zs,Seiberg:2010qd},
and even if we sum up the topological sectors
with the instanton number that is an integral multiple of~$p$,
the locality, i.e., the cluster decomposition, of the theory is not violated.
More concretely, to realize the modification of the instanton sum,
we introduce the $U(1)$ $3$-form gauge field $C^{(3)}$
and the compact scalar field $\chi$ which is the Lagrange multiplier.
As a result of the modification,
it turns out that the $\mathbb{Z}_p$ $3$-form symmetry arises.
The $4$-group structure in the YM theory under the instanton-sum modification\footnote{%
More rigorously speaking, as shown in Ref.~\cite{Hidaka:2021kkf},
this $4$-group structure should be called a semi-strict $4$-group or $3$-crossed module
in mathematical contexts.}
consists of the $\mathbb{Z}_N$ 1-form symmetry and the $\mathbb{Z}_p$ 3-form symmetry.

On the other hand, it is obvious that a continuum QFT is not well-defined
without any regularization if it does not possess triviality,
scale invariance, and so on.
For studying gauge theories, for instance,
it is important to understand the QFTs in a fully regularized framework.
We already know from L\"uscher~\cite{Luscher:1981zq}
how the $SU(N)$ principal bundle on a four-dimensional torus~$T^4$ can be constructed
in the $SU(N)$ lattice gauge theory.
Hence, for the purpose of grasping the higher-form symmetries and also higher-groups
on the basis of the lattice regularization,
it is natural to generalize L\"uscher's method
so that discrete symmetries (e.g., the center symmetry) are gauged,
such as the $\mathbb{Z}_N$ $1$-form gauge field
in the $SU(N)/\mathbb{Z}_N$ gauge theory.
Recently, a nontrivial $U(1)/\mathbb{Z}_q$ principal bundle on~$T^4$
in the compact $U(1)$ lattice gauge theory was constructed
in Ref.~\cite{Abe:2022nfq}.
In particular, due to the \emph{gauged} $\mathbb{Z}_q$ $1$-form discrete symmetry,
the $U(1)/\mathbb{Z}_q$ gauge theory exhibits the fractional topological charge
with $\frac{1}{q^2}\mathbb{Z}$ even on the finite lattice.

In this paper, we construct a higher-group structure
in the completely regularized framework, lattice regularization.
As a theory with a higher-group structure,
we consider the $U(1)$ gauge theory on~$T^4$ with restricted topological sectors
(in the same sense as the modification of the instanton sum above),
which are characterized by the integer $p$.
Similarly to the $SU(N)$ YM theory, it is shown in the continuum and the lattice that
the $U(1)$ gauge theory with this instanton-sum modification has
the $\mathbb{Z}_q$ $1$-form\footnote{%
Strictly speaking, the $U(1)$ gauge theory has
$U(1)_{\rm E} \times U(1)_{\rm M}$ 1-form symmetry,
where the former and latter are the electric and magnetic ones, respectively.
We now consider only the discrete subgroup
$\mathbb{Z}_q\times\{1\}\subset U(1)_{\rm E}\times U(1)_{\rm M}$,
where $q$ is a positive integer.}
and $\mathbb{Z}_{p q}$ $3$-form symmetries, and they also form the 4-group structure.
Because of the $4$-group structure,
we cannot gauge the $\mathbb{Z}_q$ $1$-form symmetry in a nontrivial way
without the $\mathbb{Z}_{p q}$ $3$-form \emph{gauge} symmetry.

\section{Instanton-sum modified $U(1)$ gauge theory: continuum theory}
\label{sec:mod_qed_cont}
We formulate the $U(1)$ gauge theory with the instanton-sum modification
in the continuum spacetime,\footnote{%
In this paper,
we assume that the spacetime manifold is equipped with the spin structure.}
following Ref.~\cite{Tanizaki:2019rbk}.
The modified Lagrangian of the $U(1)$ gauge theory is given by
\begin{align}
 \mathcal{L} = \frac{1}{2g^2} \mathcal{F}\wedge\star \mathcal{F}
 + \frac{i\theta}{8\pi^2} \mathcal{F}\wedge\mathcal{F}
 + i q\chi \left(\frac{1}{8\pi^2}\mathcal{F}\wedge\mathcal{F}
 - \frac{p}{2\pi}d C^{(3)}\right)
 + \frac{i\Hat{\theta}}{2\pi} d C^{(3)}.
 \label{eq:lagrangian_cont}
\end{align}
Here, $\mathcal{F}=d\mathcal{A}$ is
the field strength of the $U(1)$ gauge field $\mathcal{A}$.
By adding the third term,
the instanton number is restricted to an integral multiple of~$p$.
To see this, consider the equation of motion for the compact scalar $\chi$:
\begin{align}
 \frac{1}{8\pi^{2}}\mathcal{F}\wedge\mathcal{F}=\frac{p}{2\pi}d C^{(3)}.
 \label{eq:original_chi_eom}
\end{align}
This equation implies that topological sectors with the instanton number
that are multiples of $p$ contribute to the partition function.
Substituting Eq.~\eqref{eq:original_chi_eom} into the second term
in the Lagrangian \eqref{eq:lagrangian_cont}, we obtain the following term:
\begin{align}
 \frac{i(\Hat{\theta}+p\theta)}{2\pi}d C^{(3)}.    
\end{align}
Then, it turns out that the two theta angles are combined
into one parameter $\Hat{\theta}+p\theta$.
We can assume that $\Hat{\theta} = p \theta$, and then
the $2\pi$-periodicity of~$\Hat{\theta}$
implies that the periodicity of~$\theta$ becomes $2\pi/p$ in this instanton sector.

The usual $U(1)$ gauge theory without matter fields has
$U(1)$ $1$-form global symmetry~\cite{Gaiotto:2014kfa},
called electric symmetry $U(1)_{\rm E}$.
The gauge-invariant charged objects under the symmetry are Wilson loops.
For our purpose, we focus on two symmetries:
the $\mathbb{Z}_{q}$ $1$-form symmetry, which is a subgroup of $U(1)_{\rm E}$,
and the $\mathbb{Z}_{p q}$ $3$-form symmetry.
In order to confirm the existence of the $\mathbb{Z}_{p q}$ $3$-form global symmetry,
let us consider the equation of motion for the $3$-form gauge field $C^{(3)}$,
\begin{align}
d\chi=0.
\end{align}
It turns out from this equation that the local operator $\mathrm{e}^{i\chi}$ is
topological and generates the $3$-form symmetry.
The charged object is the volume operator,
\begin{align}
\exp\left(i\int_{M_{3}}C^{(3)}\right),
\end{align}
which is supported on a three-dimensional submanifold $M_{3}$.

Naively, it seems that these two symmetries form
a mere direct product $\mathbb{Z}_q\times\mathbb{Z}_{p q}$.
However, we find from the gauging procedure below that
they form the $4$-group structure.
As the first step of the gauging procedure,
we gauge the $\mathbb{Z}_{q}$ $1$-form symmetry alone.
Following Ref.~\cite{Kapustin:2014gua},
we introduce a $U(1)$ $2$-form gauge field $B^{(2)}$
and $U(1)$ $1$-form gauge field $B^{(1)}$ that satisfy the constraint
\begin{align}
 q B^{(2)} = d B^{(1)} .
\end{align}
In order to couple these fields to the theory,
we redefine the dynamical $U(1)$ gauge field, locally written as
\begin{align}
\mathcal{A}'=\mathcal{A}+\frac{1}{q}B^{(1)}.
\end{align}
Then, we postulate the symmetry under the $1$-form gauge transformation,
\begin{align}
 B^{(2)} &\mapsto B^{(2)} + d\Lambda^{(1)} ,&
 B^{(1)} &\mapsto B^{(1)} + q\Lambda^{(1)} ,
 \label{eq:b_shift}
\end{align}
where $\Lambda^{(1)}$ is a $U(1)$ gauge field.
Note that the gauge field $\mathcal{A}'$ also shifts as
\begin{align}
 \mathcal{A}' \mapsto \mathcal{A}' + \Lambda^{(1)}
\label{eq:a_shift}
\end{align}
by this transformation.
To ensure the gauge invariance,
the field strength $\mathcal{F}'=d\mathcal{A}'$ must be replaced by
\begin{align}
 \mathcal{F}' \to \mathcal{F}' - B^{(2)} .
\end{align}

In order to maintain the periodicity of~$\chi$, we must impose the condition
\begin{align}
 \int
 \left[\frac{q}{8\pi^2}\left(\mathcal{F}'-B^{(2)}\right)\wedge
 \left(\mathcal{F}'-B^{(2)}\right)
 - \frac{p q}{2\pi} d C^{(3)}
 \right] \in\mathbb{Z}.
 \label{eq:chi_eom}
\end{align}
The $(\mathcal{F}'\wedge\mathcal{F}')$ and $d C^{(3)}$ terms give rise to integers,
while, in general, the contribution from the quadratic term of~$B^{(2)}$ can take a value
in $\frac{1}{q}\mathbb{Z}$.\footnote{%
Recall that, in the $SU(N)/\mathbb{Z}_q$ gauge theory,
the topological charge becomes $\frac{N}{q^2}\mathbb{Z}$.}
To be consistent with the above constraint, we see that $B^{(2)}$ ($B^{(1)}$)
is restricted to a trivial configuration to give an integral value.
So we conclude that by this procedure,
we cannot gauge the $1$-form symmetry alone in a nontrivial way.

We can avoid this obstruction
by gauging the $\mathbb{Z}_{p q}$ $3$-form symmetry at the same time
as well as the discussion for the~$SU(N)$ YM theory in~Ref.~\cite{Tanizaki:2019rbk}.
We then introduce a $U(1)$ $4$-form gauge field $D^{(4)}$
and $U(1)$ $3$-form gauge field $D^{(3)}$.
Note that these fields are constrained as
\begin{align}
 p q D^{(4)} = d D^{(3)}
 - \frac{q}{4\pi} B^{(2)}\wedge B^{(2)}.
 \label{eq:def_d4}
\end{align}
As we will soon see, the above obstruction can be avoided
by imposing this constraint.
For our purpose, we redefine the $3$-form gauge field as
\begin{align}
 d C^{(3)}-D^{(4)},
\end{align}
and postulate the invariance under the $3$-form gauge transformation,
\begin{align}
 D^{(4)} &\mapsto D^{(4)} + d\Lambda^{(3)} ,&
 D^{(3)} &\mapsto D^{(3)} + p q\Lambda^{(3)} ,\\
 C^{(3)} &\mapsto C^{(3)} + \Lambda^{(3)} .
\end{align}
To keep the constraint~\eqref{eq:def_d4},
under the $1$-form transformation, we have
\begin{align}
 D^{(3)} \mapsto D^{(3)} - \frac{q}{2\pi} B^{(2)}\wedge\Lambda^{(1)}
 + \frac{q}{4\pi}\Lambda^{(1)}\wedge d\Lambda^{(1)}.
\end{align}
We can check that the constraint on $D^{(4)}$~\eqref{eq:def_d4} is invariant,
using the transformation of~$B^{(2)}$ and $B^{(1)}$ in Eq.~\eqref{eq:b_shift}
and the Bianchi identity $d\mathcal{F}'=0$.
Furthermore, the $2\pi$~periodicity of~$D^{(3)}$ is not violated by this gauge transformation.
After this gauging procedure,
we define the third term in the Lagrangian~\eqref{eq:lagrangian_cont} as
\begin{align}
i q\chi\left[\frac{1}{8\pi^{2}}\left(\mathcal{F}'-B^{(2)}\right)\wedge\left(\mathcal{F}'-B^{(2)}\right)
-\frac{p}{2\pi}dC^{(3)}+\frac{p}{2\pi}D^{(4)}\right].
\end{align}
Then, the condition~\eqref{eq:chi_eom} becomes
\begin{align}
 &\int
 \left[\frac{q}{8\pi^2}
 \left(\mathcal{F}'-B^{(2)}\right)\wedge\left(\mathcal{F}'-B^{(2)}\right)
 - \frac{p q}{2\pi} d C^{(3)} + \frac{p q}{2\pi}D^{(4)}\right] \notag\\
 &= \int \left[\frac{q}{8\pi^2}\mathcal{F}'\wedge\mathcal{F}'
 - \frac{q}{4\pi^2}\mathcal{F}'\wedge B^{(2)}
 - \frac{p q}{2\pi}d C^{(3)}+\frac{1}{2\pi}d D^{(3)}\right]
 \in\mathbb{Z} ,
 \label{eq:chi_eom_d3}
\end{align}
which is invariant under the $1$-form and $3$-form gauge transformations.
We have thus coupled the nontrivial configuration of $B^{(2)}$,
that is, the nontrivial 't~Hooft flux~\cite{tHooft:1979rtg}.
Actually, the above constraint~\eqref{eq:def_d4} provides a $\mathbb{Z}_{p q^2}$
(not $\mathbb{Z}_{p q}$) phase:
\begin{align}
 \int D^{(4)}=\frac{2\pi}{p q^{2}}
 \left(
  q\int\frac{dD^{(3)}}{2\pi}
  +q\int\frac{\mathcal{F}'}{2\pi}\wedge\frac{qB^{(2)}}{2\pi}
  -\frac{1}{2}\int\frac{qB^{(2)}}{2\pi}\wedge\frac{qB^{(2)}}{2\pi}
 \right)
 \in\frac{2\pi}{p q^{2}}\mathbb{Z}.
\end{align}
Hence,
one can obtain the extra phase factor under the shift~$\theta\to\theta+2\pi/p$
or $\Hat{\theta}\to\Hat{\theta}+2\pi$,
\begin{align}
 Z_{\theta+2\pi/p} = e^{-i\int D^{(4)}} Z_{\theta} .
\end{align}
This construction means that the $U(1)$ gauge theory
under the instanton-sum modification
possesses the $4$-group structure between the $1$-form and $3$-form symmetries.

In the usual statement of the higher-group structure,
$D^{(4)}$ is not invariant under the $1$-form transformation
and is also a $\mathbb{Z}_{p q}$ $4$-form gauge field.
On the other hand, our construction indicates that $D^{(4)}$ is the $\mathbb{Z}_{p q^2}$ gauge field.
One may define a genuine $\mathbb{Z}_{p q}$ gauge field $\mathcal{D}^{(4)}$ by
\begin{align}
 p q\mathcal{D}^{(4)} = d D^{(3)} ,
\end{align}
instead of~$D^{(4)}$.
Then $\mathcal{D}^{(4)}$ is not $1$-form-invariant as
\begin{align}
 p q\mathcal{D}^{(4)} \mapsto p q\mathcal{D}^{(4)}
 + \frac{q}{2\pi}B^{(2)}\wedge d\Lambda^{(1)}
 + \frac{q}{4\pi}d\Lambda^{(1)}\wedge d\Lambda^{(1)} .
\end{align}
In this case, we also have the same equation of motion~\eqref{eq:chi_eom_d3}.

\section{Setup of lattice $U(1)/\mathbb{Z}_q$ gauge theory}
\subsection{Lattice construction of the $U(1)/\mathbb{Z}_q$ principal bundle}
Let us now review the formulation of the lattice $U(1)/\mathbb{Z}_q$ gauge theory
defined in~Ref.~\cite{Abe:2022nfq}.
We consider a four-dimensional periodic torus~$T^4$ and its discretization~$\Lambda$,
\begin{align}
 T^4&\equiv\left\{x\in\mathbb{R}^4 \mid \text{$0\leq x_\mu<L$ for all $\mu$}\right\} ,
 \\
 \Lambda&\equiv\left\{n\in\mathbb{Z}^4
 \mid \text{$0\leq n_\mu<L$ for all $\mu$}\right\} ,
\end{align}
where $L$ is the size, the lattice constant is set to $1$,
and $\mu$ (also $\nu$, $\rho$, \dots) runs over $1$, $2$, $3$, and $4$.
This lattice~$\Lambda$ divides $T^4$ into hypercubes~$c(n)$ as
\begin{align}
 c(n)\equiv\left\{x\in\mathbb{R}^4
 \mid \text{$0\leq(x_\mu-n_\mu)\leq1$ for all $\mu$}\right\} .
\end{align}
Now, $c(n)$ denotes a patch of the fiber bundle under consideration.
Then the transition function \emph{on the lattice}
is defined at the boundary of two hypercubes, called the face,
\begin{align}
 f(n,\mu)\equiv\left\{x\in c(n) \mid x_\mu=n_\mu\right\} = c(n-\Hat{\mu})\cap c(n),
\end{align}
where $\Hat{\mu}$ is a unit vector in the positive $\mu$ direction.

Now we assume the lattice $U(1)$ gauge theory with the link variable
\begin{align}
 U(n,\mu)\in U(1),
\end{align}
which lives on the link connecting $n$ and~$n+\Hat{\mu}$.
The field strength is defined by
\begin{align}
 \Check{F}_{\mu\nu}(n)
 \equiv \frac{1}{i q}\ln\left[U(n,\mu)U(n+\Hat{\mu},\nu)
 U(n+\Hat{\nu},\mu)^{-1}U(n,\nu)^{-1}\right]^q
 \qquad -\pi<\Check{F}_{\mu\nu}(n)\leq\pi.
\end{align}
Here, we require the above dependence on the integer~$q$
because of the $\mathbb{Z}_q$ $1$-form gauge symmetry we will define below.
Following L\"uscher's construction of the $SU(N)$ principal bundle
in~$\Lambda$~\cite{Luscher:1981zq},
we have the simpler transition function~$\Check{v}_{n,\mu}(x)$
at $x\in f(n,\mu)$ for the $U(1)$ gauge group.\footnote{%
The expressions are somewhat long to write it here but not essential in this paper;
for the explicit expressions, see Refs.~\cite{Fujiwara:2000wn,Abe:2022nfq}.}
Under ordinary circumstances of the fiber bundle,
the cocycle condition for the transition function
is given at the boundary of three patches,
but on the lattice, it can be defined in a two-dimensional plaquette,
that is, the intersection of four hypercubes,
$c(n)$, $c(n-\Hat{\mu})$, $c(n-\Hat{\nu})$, and~$c(n-\Hat{\mu}-\Hat{\nu})$:
\begin{align}
 p(n,\mu,\nu)\equiv\left\{x\in c(n) \mid \text{$x_\mu=n_\mu$, $x_\nu=n_\nu$}\right\}
 \qquad (\mu\neq\nu).
\end{align}
Then, for~$x\in p(n,\mu,\nu)$, the transition function satisfies the cocycle condition,
\begin{align}
 \Check{v}_{n-\Hat{\nu},\mu}(x)\Check{v}_{n,\nu}(x)
 \Check{v}_{n,\mu}(x)^{-1}\Check{v}_{n-\Hat{\mu},\nu}(x)^{-1} = 1.
 \label{eq:cocycle_check}
\end{align}

From now on, we consider the $U(1)/\mathbb{Z}_q$ principal bundle.
Then, the cocycle condition~\eqref{eq:cocycle_check}
can be relaxed to be $\mathbb{Z}_q$-valued.
Following the construction in Ref.~\cite{Abe:2022nfq},
in terms of~$\Check{v}_{n,\mu}(x)$,
we can define the transition function for~$U(1)/\mathbb{Z}_q$, $v_{n,\mu}(x)$,
at~$x\in f(n,\mu)$ as follows:
We first introduce the 't~Hooft flux at~$x_\mu=0$,
so we define the transition function initially as
\begin{align}
 v_{n,\mu}(x)\equiv\omega_\mu(x)\Check{v}_{n,\mu}(x) ,
\end{align}
where we have introduced the loop factor~$\omega_\mu(x)$ as
\begin{align}
 \omega_\mu(x)\equiv
 \begin{cases}
  \exp\left(\frac{\pi i}{q}\sum_{\nu\neq\mu}\frac{z_{\mu\nu} x_\nu}{L}\right)
  & \text{for $x_\mu=0\bmod L$},\\
  1 & \text{otherwise}
 \end{cases}
\end{align}
and $z_{\mu\nu}\in\mathbb{Z}$ is anti-symmetric in indices, $z_{\nu\mu}=-z_{\mu\nu}$.
Next we consider the~$\mathbb{Z}_q$ $1$-form gauge transformation given by
\begin{align}
 U(n,\mu)&\mapsto\exp\left[\frac{2\pi i}{q}z_\mu(n)\right]U(n,\mu) ,\\
 v_{n,\mu}(x)&\mapsto\exp\left[\frac{2\pi i}{q}z_\mu(n-\Hat{\mu})\right]v_{n,\mu}(x)
 \qquad \text{at $x\in f(n,\mu)$},
\end{align}
where we have used $z_\mu(n)\in\mathbb{Z}$, and
\begin{align}
 0\leq z_\mu(n)<q \qquad \text{for $\forall \mu$}.
\end{align}
Then, in terms of $z_{\mu\nu}$
and $1$-form fields $\{z_\mu(n)\}$ for successive transformations,
we have a generic $\mathbb{Z}_q$ $2$-form gauge field~$z_{\mu\nu}(n)$,
and the transition function~$v_{n,\mu}(x)$ associated with~$z_{\mu\nu}(n)$.
The cocycle condition is modified to
\begin{align}
 v_{n-\Hat{\nu},\mu}(x)v_{n,\nu}(x)v_{n,\mu}(x)^{-1}v_{n-\Hat{\mu},\nu}(x)^{-1}
 = \exp\left[\frac{2\pi i}{q}z_{\mu\nu}(n-\Hat{\mu}-\Hat{\nu})\right]\in\mathbb{Z}_q .
\end{align}
Under the $\mathbb{Z}_q$ $1$-form gauge transformation, we find
\begin{align}
 z_{\mu\nu}(n)
 \mapsto z_{\mu\nu}(n) + \Delta_\mu z_\nu(n) - \Delta_\nu z_\mu(n) + q N_{\mu\nu}(n) ,
 \label{eq:1_transf}
\end{align}
where we have defined the forward difference,
$\Delta_\mu f(n)\equiv f(n+\Hat{\mu})-f(n)$.
Note that there exists the ambiguity of~$z_{\mu\nu}(n) \bmod q$,
and thus the field~$N_{\mu\nu}(n)\in\mathbb{Z}$ is demanded to restrict
$z_{\mu\nu}(n)$ to a single-valued lattice field, for instance, as
\begin{align}
 \begin{cases}
  0\leq z_{\mu\nu}(n)<q & \text{for $\mu<\nu$},\\
  z_{\mu\nu}(n)\equiv - z_{\nu\mu}(n) & \text{for $\mu>\nu$}.
 \end{cases}
 \label{eq:z_range}
\end{align}

\subsection{Topological charge in terms of non-commutative differential calculus}
We will show the expression of the fractional topological charge
in the $U(1)/\mathbb{Z}_q$ principal bundle.
In comparison with continuum theory,
it may be convenient to use the non-commutative differential calculus~\cite{Dimakis:1992pk}.
As usual, we define a $k$-form $f(n)$ on the lattice by
\begin{align}
 f(n)\equiv\frac{1}{k!}\sum_{\mu_1,\dots,\mu_k} f_{\mu_1\dots\mu_k}(n)
 d x_{\mu_1}\wedge\dots\wedge d x_{\mu_k} ,
\end{align}
and the exterior derivative on~$\Lambda$ by
\begin{align}
 d f(n)\equiv\frac{1}{k!}\sum_{\mu,\mu_1,\dots,\mu_k}
 \Delta_\mu f_{\mu_1\dots\mu_k}(n)
 d x_\mu\wedge d x_{\mu_1}\wedge \dots\wedge d x_{\mu_k}.
\end{align}
Note that $d^2=0$ is satisfied (nilpotent).
The essential formal rule is now given by
\begin{align}
 d x_\mu\wedge f(n) = (-1)^k f(n+\Hat{\mu})\wedge d x_\mu ,
\end{align}
and then the Leibniz rule holds even on the finite lattice,
\begin{align}
 d\left[f(n)\wedge g(n)\right]
 = d f(n)\wedge g(n) + (-1)^k f(n)\wedge d g(n) .
\end{align}

Now we rewrite the field strength and the $\mathbb{Z}_q$ $2$-form gauge field by
\begin{align}
 \Check{F}(n) &\equiv \frac{1}{2}\sum_{\mu,\nu}
 \Check{F}_{\mu\nu}(n)d x_\mu\wedge d x_\nu,&
 z^{(2)}(n) &\equiv \frac{1}{2}\sum_{\mu,\nu}z_{\mu\nu}(n)d x_\mu\wedge d x_\nu,
\end{align}
and also
\begin{align}
 z^{(1)}(n) &\equiv \sum_{\mu}z_\mu(n)d x_\mu ,&
 N^{(2)}(n) &\equiv \frac{1}{2}\sum_{\mu,\nu}N_{\mu\nu}(n)d x_\mu\wedge d x_\nu .
\end{align}
The $\mathbb{Z}_q$ $1$-form gauge transformation~\eqref{eq:1_transf} is rewritten as
\begin{align}
 z^{(2)}(n)\mapsto z^{(2)}(n) + d z^{(1)}(n) + q N^{(2)}(n) .
 \label{eq:z_transf}
\end{align}
For an \emph{admissible} gauge field,
the Bianchi identity, $d\Check{F}(n)=0$, should hold;
we also find that the flatness condition, $d z^{(2)}(n)=0\bmod q$.
In what follows, we use the abbreviated notations,
$z^{(2)}(n)\in\mathbb{Z}$, which indicates that $z_{\mu\nu}(n)\in\mathbb{Z}$ and so on;
$0\leq z^{(2)}(n)<q$ denotes Eq.~\eqref{eq:z_range},
and also, if we say $0\leq f(n)<q$ for a generic $k$-form $f(n)$,
we mean that $0\leq f_{\mu_1\dots\mu_k}<q$ for $\mu_1<\dots<\mu_k$.

With knowledge of the continuum theory, the topological charge
$Q = \frac{1}{8\pi^2} \int_{T^4}\mathcal{F}\wedge\mathcal{F}$
can be written, in terms of the transition function,
by~\cite{Luscher:1981zq,vanBaal:1982ag}
\begin{align}
 Q = -\frac{1}{8\pi^2}\sum_{n\in\Lambda}\sum_{\mu,\nu,\rho,\sigma}
 \epsilon_{\mu\nu\rho\sigma} \int_{p(n,\mu,\nu)} d^2 x
 \left[v_{n,\mu}(x)\partial_\rho v_{n,\mu}(x)^{-1}\right]
 \left[v_{n-\Hat{\mu},\nu}(x)^{-1}\partial_\sigma v_{n-\Hat{\mu},\nu}(x)\right] .
\end{align}
Thus, one finds that~\cite{Abe:2022nfq}
\begin{align}
 Q = \frac{1}{8\pi^2}\sum_{n\in\Lambda}\left[F(n)+\frac{2\pi}{q}z^{(2)}(n)\right]
 \wedge\left[F(n)+\frac{2\pi}{q}z^{(2)}(n)\right] ,
 \label{eq:q_lat}
\end{align}
where this expression on the lattice is similar to that in the continuum theory
up to the possible definitions of the $\mathbb{Z}_q$ $2$-form gauge fields,
$z^{(2)}$ and $B^{(2)}$, respectively.
Here, another field strength $F(n)$ can be constructed from~$\Check{F}(n)$
by the procedure of successive transformations in the previous subsection,
supposing that $F(n)$ is not invariant
under the $\mathbb{Z}_q$ $1$-form transformation as
\begin{align}
 F(n)\mapsto F(n)-\frac{2\pi}{q}\left[d z^{(1)}(n) + q N^{(2)}(n)\right] .
\end{align}
This is analogous to~Eq.~\eqref{eq:a_shift}, that is,
the shift of $\mathcal{A}'$ in the continuum theory under the $1$-form transformation.
From this expression~\eqref{eq:q_lat},
we see the fractionality of the topological charge
because of the coefficient~$1/q$ in the $z^{(2)}(n)$ terms.

One may wonder that the above expression~\eqref{eq:q_lat}
implies $Q\in\frac{1}{2q^2}\mathbb{Z}$.
The topological charge naively seems to suffer from
the non-commutativity in $z^{(2)}(n)\wedge z^{(2)}(n)
=(1/4)\sum_{\mu,\nu,\rho,\sigma}\epsilon_{\mu\nu\rho\sigma}z_{\mu\nu}(n)z_{\rho\sigma}(n+\Hat{\mu}+\Hat{\nu})d^4 x$.
This is, in fact, not harmful, because the flatness condition, $d z^{(2)}(n)=0\bmod q$,
that leads us to the shift of~$\sum_{n\in\Lambda}z^{(2)}(n)\wedge z^{(2)}(n)$
under the $1$-form gauge transformation
is $q\mathbb{Z}$-valued; if $d z^{(2)}(n)=0$ strictly, we find $2q\mathbb{Z}$.
Thus, we cannot vary the number proportional to~$1/q^2$
from the original contribution of the 't~Hooft flux,
$\frac{1}{8q^2}\sum_{\mu,\nu,\rho,\sigma}\epsilon_{\mu\nu\rho\sigma}z_{\mu\nu}z_{\rho\sigma}
\in\frac{1}{q^2}\mathbb{Z}$.
It is however obstructive that the above shift by~$q\mathbb{Z}$ would produce
terms with~$\frac{1}{2q}\mathbb{Z}$.
The $U(1)/\mathbb{Z}_q$ gauge theory is, of course, not hindered by this fact,
since the total expression of~$Q$ is $1$-form invariant.
On the other hand,
if we consider the instanton-sum modification and the higher-group structure,
the $1$-form and $3$-form gauge transformations are inseparable from each other,
and so it is crucial to have the gauge transformation shift by~$2q\mathbb{Z}$,
as we will show.

Actually, the above argument would be an artifact on the square lattice,
while the simplicial lattice does not give rise to the extra factor~$1/2$.
In a simplicial complex,
the wedge products should be replaced by the Pontryagin square, which can be constructed
by (higher) cup products~\cite{Kapustin:2014gua,Kapustin:2013qsa}.
Because of the difficulty in constructing the Pontryagin square on the square lattice,
we cannot simply write some ``finest'' local terms.

If we consider a torsion-free manifold (e.g., the torus),
we can circumvent this problem by a technique
called the integral lift~\cite{Kapustin:2014gua,Kapustin:2013qsa}.
The integral lift enables us to define $\Bar{z}^{(2)}(n)$ for $\forall z^{(2)}(n)$
such that $z^{(2)}(n)=\Bar{z}^{(2)}(n)\bmod q$ and $d\Bar{z}^{(2)}(n)=0$.
Now, $H^2(T^4,\mathbb{Z}_q)$ is ``lifted'' up to $H^2(T^4,\mathbb{Z})$.
The detailed definition on the square lattice, which seems to be technically cumbersome,
is given in Sect.~4 of~Ref.~\cite{Abe:2022nfq}.
Therefore, we have
\begin{align}
 \Bar{Q} = \frac{1}{8\pi^2}\sum_{n\in\Lambda}
 \left[F(n)+\frac{2\pi}{q}\Bar{z}^{(2)}(n)\right]
 \wedge\left[F(n)+\frac{2\pi}{q}\Bar{z}^{(2)}(n)\right] .
 \label{eq:qbar_lat}
\end{align}
Now, $\Bar{Q}$ is again invariant under the $1$-form transformation given by
\begin{align}
 \Bar{z}^{(2)}(n) &\mapsto \Bar{z}^{(2)}(n) + d z^{(1)}(n) + q\Bar{N}^{(2)}(n) ,
 \label{eq:zbar_transf}\\
 F(n) &\mapsto F(n)-\frac{2\pi}{q}\left[d z^{(1)}(n) + q\Bar{N}^{(2)}(n)\right] ,
\end{align}
and the flatness conditions are
\begin{align}
 dF(n) = d\Bar{z}^{(2)}(n) = d\Bar{N}^{(2)}(n) = 0 .
\end{align}

\section{Instanton-sum modified $U(1)$ gauge theory: lattice theory}
In this section, we consider the system given in Sect.~\ref{sec:mod_qed_cont},
but on the basis of the lattice regularization.
We again insert the delta function after the integration over~$\chi$
in the path integral on the lattice\footnote{%
It is actually quite subtle on the lattice
to give the naive coupling of the compact scalar~$\chi$
as in Eq.~\eqref{eq:lagrangian_cont}.
Thus, for our lattice formulation,
it would be unambiguous to insert the delta function
instead of the coupling of the Lagrange multiplier.
For this difficulty, e.g., see Ref.~\cite{Cordova:2019jnf}.
We would like to thank Yuya Tanizaki for pointing it out to us.}
\begin{align}
 \delta((q/8\pi^2)\Check{F}(n)\wedge\Check{F}(n)
 -p q c^{(4)}(n)) ,
 \label{eq:chi_lat}
\end{align}
where $c^{(4)}(n)$ is the field strength of a compact $U(1)$ $3$-form gauge field,
i.e., a $4$-form such that $\sum_{n\in\Lambda}c^{(4)}(n)\in\mathbb{Z}$
(not a conventional definition of the field strength
by the overall factor~$2\pi$ for simplicity below).
This restricts the instanton number to $p\mathbb{Z}$ on the $U(1)$ principal bundle.
Now, we consider the gauged $\mathbb{Z}_q$ $1$-form symmetry as in the previous section.
Then we find that the equation of motion for~$\chi$ is given by
\begin{align}
 \frac{q}{8\pi^2}\left[F(n)+\frac{2\pi}{q}\Bar{z}^{(2)}(n)\right]\wedge
 \left[F(n)+\frac{2\pi}{q}\Bar{z}^{(2)}(n)\right]
 - p q c^{(4)}(n) = 0.
\end{align}
As in the case in the continuum, at this stage,
the nontrivial configurations of the $2$-form gauge field $\Bar{z}^{(2)}$ are forbidden.

To gauge the $1$-form symmetry,
the fractional parts of the above equation of motion can be compensated by
inserting new $4$-form field~$\Bar{w}^{(4)}(n)\in\mathbb{Z}$ by hand,
\begin{align}
 \frac{q}{8\pi^2}\left[F(n)+\frac{2\pi}{q}\Bar{z}^{(2)}(n)\right]\wedge
 \left[F(n)+\frac{2\pi}{q}\Bar{z}^{(2)}(n)\right]
 - p q c^{(4)}(n) + \frac{1}{q} \Bar{w}^{(4)}(n) = 0 .
\end{align}
We now see that all fractional parts in $\frac{1}{q}\mathbb{Z}$
can be absorbed into the last term,
$\frac{1}{q}\Bar{w}^{(4)}(n)\in\frac{1}{q}\mathbb{Z}$,
and thus the allowable gauge field~$\Bar{z}^{(2)}$ is
what we have wanted in order to gauge the $\mathbb{Z}_q$ $1$-form symmetry.
From this compensation procedure, we come to realize that
$c^{(4)}(n)$ could be replaced systematically
by~$c^{(4)}\to c^{(4)}-\frac{1}{p q^2}\Bar{w}^{(4)}(n)$.
This fact indicates that we have the gauged $3$-form shift symmetry,
\begin{align}
 \Bar{w}^{(4)}(n) &\mapsto \Bar{w}^{(4)}(n) + d w^{(3)}(n) + p q^2 \Bar{N}^{(4)}(n) ,
 \label{eq:w4_transf}\\
 c^{(4)}(n) &\mapsto c^{(4)}(n) + \frac{1}{p q^2} d w^{(3)}(n) + \Bar{N}^{(4)}(n) .
 \label{eq:c4_transf}
\end{align}
Here, we assume that $w^{(3)}(n)\in\mathbb{Z}$ satisfies
$\sum_{n\in\Lambda}d w^{(3)}(n)\in p q^2\mathbb{Z}$
and $0\leq w^{(3)}(n)<p q^2$;
$w^{(3)}(n)$ and $\Bar{N}^{(4)}(n)\in\mathbb{Z}$ have been introduced
to give the fractional and integral shifts for~$c^{(4)}(n)$, respectively.
It would seem to accomplish our task to gauge
the $\mathbb{Z}_q$ $1$-form and $3$-form symmetries.

There is a question we left unanswered:
what discrete group is induced in the above $3$-form gauge symmetry
represented by the $4$-form gauge field~$\Bar{w}^{(4)}(n)$?
We now recall that the $\mathbb{Z}_q$ $1$-form gauge transformation of~$z^{(2)}(n)$
is given by Eq.~\eqref{eq:z_transf},
and it is very similar to the $3$-form gauge transformation
of~$\Bar{w}^{(4)}(n)$ \eqref{eq:w4_transf}.
Roughly speaking, this observation and the restriction~$0\leq w^{(3)}(n)<p q^2$
could imply that $\Bar{w}^{(4)}(n)$ is the $\mathbb{Z}_{p q^2}$ $4$-form gauge field,
while the original $3$-form symmetry looks like $\mathbb{Z}_{p q}$
if we ignore $\Bar{z}^{(2)}(n)$ and the dependence on~$q$.
Hence, this is the same $4$-group structure
as the $\mathbb{Z}_{p q^2}$ $4$-form gauge field $D^{(4)}$ in the continuum.

Of course, the above argument would be true but so naive,
and the situation is more complicated
because $\Bar{w}^{(4)}(n)$ seems to be leading
not to a $\mathbb{Z}_{p q^2}$ gauge field but $\mathbb{Z}$.
As we already know, this is just the integral lift.
Conversely,
we now redefine $\Bar{w}^{(4)}(n)$ as $w^{(4)}(n)=\Bar{w}^{(4)}(n)\bmod p q^2$,
\begin{align}
 w^{(4)}(n) = w(n) d x_1\wedge d x_2\wedge d x_3\wedge d x_4,
 \qquad
 w(n)\in\mathbb{Z},\ 
 0\leq w(n) < p q^2 .
\end{align}
Also, $N^{(4)}(n)$ instead of~$\Bar{N}^{(4)}(n)$ in Eq.~\eqref{eq:w4_transf}
should take a value to keep $w^{(4)}(n)$ in this region, depending on~$w^{(3)}(n)$.
We should note that, on the other hand,
in Eqs.~\eqref{eq:w4_transf} and~\eqref{eq:c4_transf},
the lifted field $\Bar{N}^{(4)}(n)$ can take any integers.
In fact, $\Bar{N}^{(2)}(n)$ is determined
from the amount of the ``relaxation'' of flatness, $d z^{(2)}(n)/q\in\mathbb{Z}$.
However, since any $4$-form is closed,
we have many degrees of freedom to fix the lifted field~$\Bar{N}^{(4)}(n)$.
Moreover, the role of~$N^{(4)}(n)$ or~$\Bar{N}^{(4)}(n)$
in the $3$-form transformation of~$c^{(4)}(n)$
is almost trivial due to the definition of~$c^{(4)}(n)$; see Eq.~\eqref{eq:chi_lat} below.
Our argument successfully provides
the description of the higher-group structure on the lattice.

In the above proposition to absorb the fractional terms
in the equation of motion for~$\chi$,
one might promote the use of~$\Bar{w}^{(4)}(n)\in\mathbb{Z}$
to~$\Omega^{(4)}(n)\in\mathbb{R}$.
We then again have
\begin{align}
 \frac{q}{8\pi^2}\left[F(n)+\frac{2\pi}{q}\Bar{z}^{(2)}(n)\right]\wedge
 \left[F(n)+\frac{2\pi}{q}\Bar{z}^{(2)}(n)\right]
 - p q c^{(4)}(n) + \frac{1}{q}\Omega^{(4)}(n) = 0,
\end{align}
where we assume that $\sum_{n\in\Lambda}\Omega^{(4)}(n)\in\mathbb{Z}$
to get rid of the fractionality in this equation.
We now define $\Tilde\Omega^{(4)}(n)\in\mathbb{R}$ as
\begin{align}
 \Tilde\Omega^{(4)}(n) &\equiv \frac{1}{q} \Omega^{(4)}(n)
 + \frac{1}{4\pi} \left[F(n)\wedge\Bar{z}^{(2)}(n)+\Bar{z}^{(2)}(n)\wedge F(n)\right]
 \notag\\&\qquad
 + \frac{1}{2q} \Bar{z}^{(2)}(n)\wedge\Bar{z}^{(2)}(n),
\end{align}
and therefore, we see $\sum_{n\in\Lambda}\Tilde\Omega^{(4)}(n)\in\mathbb{Z}$
and the equation of motion
\begin{align}
 \frac{q}{8\pi^2} F(n)\wedge F(n) - p q c^{(4)}(n) + \Tilde\Omega^{(4)}(n) = 0.
\end{align}
We have seen that $w^{(4)}(n)$ is invariant
under the $\mathbb{Z}_q$ $1$-form gauge transformation,
while we now find that
\begin{align}
 \Tilde\Omega^{(4)}(n) &\mapsto \Tilde\Omega^{(4)}(n)
\notag\\&\qquad{}
 + \frac{1}{4\pi}F(n)\wedge\left[d z^{(1)}(n) + q\Bar{N}^{(2)}(n)\right]
 + \frac{1}{4\pi}\left[d z^{(1)}(n) + q\Bar{N}^{(2)}(n)\right]\wedge F(n)
\notag\\&\qquad\qquad{}
 - \frac{1}{2q} \left[d z^{(1)}(n) + q\Bar{N}^{(2)}(n)\right]\wedge
 \left[d z^{(1)}(n) + q\Bar{N}^{(2)}(n)\right] .
 \label{eq:omega_transf1}
\end{align}
It is interesting that, noting that
the shift of~$\sum_{n\in\Lambda}\Bar{z}^{(2)}(n)\wedge\Bar{z}^{(2)}(n)$
under the $1$-form gauge transformation is $2q\mathbb{Z}$-valued,
we can see that $\sum_{n\in\Lambda}\Tilde\Omega^{(4)}(n)\in\mathbb{Z}$ holds
even if we perform Eq.~\eqref{eq:omega_transf1}.
The $3$-form gauge transformation is also given by
\begin{align}
 \Tilde\Omega^{(4)}(n) &\mapsto \Tilde\Omega^{(4)}(n)
 + d\Tilde\Omega^{(3)}(n) + p q\Tilde{N}^{(4)}(n) ,
 \label{eq:omega_transf3}\\
 c^{(4)}(n) &\mapsto c^{(4)}(n) + \frac{1}{p q}d\Tilde\Omega^{(3)}(n)
 + \Tilde{N}^{(4)}(n) ,
\end{align}
where $\Tilde\Omega^{(3)}(n)\in\mathbb{R}$ and $\Tilde{N}^{(4)}(n)\in\mathbb{Z}$
give rise to the shift of~$c^{(4)}(n)$ by real and integer numbers.

We now suppose that $\Tilde\Omega^{(4)}(n)\in\mathbb{R}$
is compelled to be $\mathbb{Z}$.
To this end, there are two steps as follows:
(i) The $3$-form transformation~\eqref{eq:omega_transf3} is restricted to that
with the parameter~$\Tilde\Omega^{(3)}(n)\to\Tilde{w}^{(3)}(n)\in\mathbb{Z}$.
So we call this the $3$-form discrete gauge transformation,
while the original one is the continuum transformation;
(ii) First, we consider the $\mathbb{Z}_q$ $1$-form gauge transformation
given by~\eqref{eq:omega_transf1},
but then $\Tilde\Omega^{(4)}(n)\notin\mathbb{Z}$
even if we start with~$\Tilde\Omega^{(4)}(n)\in\mathbb{Z}$.
We next impose that $\Tilde\Omega^{(4)}(n)$ becomes
a $\mathbb{Z}$-valued field $\Tilde{\Bar{w}}^{(4)}(n)$.
To do this,
we use the $3$-form continuum gauge transformation by~$\Tilde\Omega^{(3)}(n)$,
that is, throw away real parts apart from integers in~$\Tilde\Omega^{(4)}(n)$
into the gauge redundancy of $c^{(4)}(n)$.
Eventually, we define $\Tilde{\Bar{w}}^{(4)}(n)\in\mathbb{Z}$
as the integral part of~$\Tilde\Omega^{(4)}(n)$.

From the above construction of~$\Tilde{\Bar{w}}^{(4)}(n)$,
we can see the $3$-form gauge transformation
\begin{align}
 \Tilde{\Bar{w}}^{(4)}(n) &\mapsto \Tilde{\Bar{w}}^{(4)}(n)
 + d\Tilde{w}^{(3)}(n) + p q\Tilde{\Bar{N}}^{(4)}(n) ,
\end{align}
where $\Tilde{w}^{(3)}(n)$ is now an element in $\mathbb{Z}$.
Our argument about the integral lift is again
applicable to this $3$-form gauge symmetry,
so $\Tilde{w}^{(4)}(n)\equiv\Tilde{\Bar{w}}^{(4)}(n)\bmod p$ denotes
the $\mathbb{Z}_{p q}$ $4$-form gauge field.
Thus, we have the $\mathbb{Z}_q$ $1$-form
and $\mathbb{Z}_{p q}$ $3$-form discrete gauge symmetries,
which are inseparable from each other because of the above argument (ii).
This gives another description of the $4$-group structure
in the lattice Abelian gauge theory on the $p\mathbb{Z}$-topological sector.

\section{Conclusion}
In the present paper,
we constructed the instanton-sum modified $U(1)$ lattice gauge theory on $T^4$.
We firstly showed that the naive gauging of the $\mathbb{Z}_q$ $1$-form symmetry
is not compatible with the equation of motion for $\chi$
introduced for modification of the instanton summation in $U(1)$ gauge theory
on the continuum manifold, as in the $SU(N)$ gauge theory.
This is the same in lattice gauge theory.
In order to compensate for the equation of motion,
we see that it is necessary to introduce
the $4$-form field $\Bar{w}^{(4)}(n) \in \mathbb{Z}$.
Then the integral lift plays an important role:
the $4$-form field $\Bar{w}^{(4)}(n)$ is interpreted
as the lifted field of the $\mathbb{Z}_{p q^2}$ $4$-form gauge field $w^{(4)}(n)$
which is the gauge field of the $3$-form symmetry
and corresponds to the $\mathbb{Z}_{p q^2}$ gauge field $D^{(4)}$ in the continuum theory.
In this construction, $w^{(4)}(n)$ and its lifted field $\bar{w}^{(4)}(n)$
do not transform under the $\mathbb{Z}_q$ $1$-form symmetry.
In addition to the above construction, we also considered
introducing $\tilde{\Omega}^{(4)}(n)\in \mathbb{R}$
instead of $\bar{w}^{(4)}(n) \in \mathbb{Z}$ to compensate for the equation of motion.
Then we can recognize that the integral part of $\tilde{\Omega}^{(4)}(n)$ is
the integral lift of the $\mathbb{Z}_{p q}$ 4-form gauge field $\tilde{w}^{(4)}(n)$
corresponding to the $\mathbb{Z}_{p q}$ gauge field $\mathcal{D}^{(4)}$ in the continuum theory.
In this case, in contrast, $\tilde{w}^{(4)}(n)$ transforms
not only under the $3$-form symmetry but also under $1$-form symmetry,
such as the $1$-form gauge field of the Green--Schwarz mechanism.

It is interesting to generalize our construction
to the $SU(N)$ gauge theory with the instanton-sum modification.
To accomplish this, we need to construct
a nontrivial $SU(N)/\mathbb{Z}_N$ principal bundle in the lattice gauge theory.
It is also intriguing to consider lattice formulations with matter fields,
and especially to see the higher-group structure for the lattice fermion,
whose construction is an attractive issue in lattice gauge theory.

\section*{Acknowledgements}
We are grateful to Yuya Tanizaki, Akio Tomiya, Satoshi Yamaguchi, and Ryo Yokokura for useful discussions.
O.M. also would like to thank Motokazu Abe, Yuki Miyakawa, Soma Onoda, and Hiroshi Suzuki for collaboration and helpful discussions.
This work was partially supported by the Japan Society for the Promotion of Science (JSPS)
Grant-in-Aid for Scientific Research Grant Number JP21J30003 (O.M.)

\bibliographystyle{utphys}
\bibliography{ref}
\end{document}